\documentclass[10pt,twocolumn,pagenumbers]{article}

\usepackage[pagenumbers]{wacv} 

\definecolor{wacvblue}{rgb}{0.21,0.49,0.74}
\usepackage[pagebackref,breaklinks,colorlinks,allcolors=wacvblue]{hyperref}

\def\wacvPaperID{1664} 
\def\confName{WACV}
\def\confYear{2026}
\usepackage{tgpagella}
\usepackage{balance}

\title{MLRU++: Multiscale Lightweight Residual UNETR++ with Attention for Efficient 3D Medical Image Segmentation}
\author{$^\dagger$Nand Kumar Yadav\thanks{This work was supported in part by the National Science Foundation under Award \href{https://www.nsf.gov/awardsearch/showAward?AWD_ID=2346643}{\#2346643} and in part by the U.S. Department of Defense (DoD) under Award \href{https://dtic.dimensions.ai/details/grant/grant.14525543}{\#FA95502310495}.}, $^\dagger$Rodrigue Rizk, $^\ddagger$William CW Chen, $^\dagger$KC Santosh\\
$^\dagger$AI Research Lab, Department of Computer Science\\
$^\ddagger$Biomedical \& Translational Sciences, Sanford School of Medicine\\
University Of South Dakota, Vermillion, SD 57069 USA.\\
{\tt\small \{nand.yadav, rodrigue.rizk@usd.edu, william.chen, kc.santosh\}@usd.edu}
}

\begin{document}
\maketitle
\begin{abstract}
Accurate and efficient medical image segmentation is crucial but challenging due to anatomical variability and high computational demands on volumetric data. 
Recent hybrid CNN-Transformer architectures achieve state-of-the-art results but add significant complexity.
In this paper, we propose MLRU++, a  Multiscale Lightweight Residual UNETR++ architecture designed to balance segmentation accuracy and computational efficiency. It introduces two key innovations: a Lightweight Channel and Bottleneck Attention Module (LCBAM) that enhances contextual feature encoding with minimal overhead, and a Multiscale Bottleneck Block (M\textsuperscript{2}B) in the decoder that captures fine-grained details via multi-resolution feature aggregation. Experiments on four publicly available benchmark datasets (Synapse, BTCV, ACDC, and Decathlon Lung) demonstrate that MLRU++ achieves state-of-the-art performance, with average Dice scores of 87.57\% (Synapse), 93.00\% (ACDC), and 81.12\% (Lung). Compared to existing leading models, MLRU++ improves Dice scores by 5.38\% and 2.12\% on Synapse and ACDC, respectively, while significantly reducing parameter count and computational cost. 
Ablation studies evaluating LCBAM and M\textsuperscript{2}B further confirm the effectiveness of the proposed architectural components. Results suggest that MLRU++ offers a practical and high-performing solution for 3D medical image segmentation tasks. Source code is available at: \href{https://github.com/1027865/MLRUPP}{https://github.com/1027865/MLRUPP}.

\end{abstract}

\section{Introduction}
\label{sec:intro}
Accurate 3D medical image segmentation is crucial for a variety of reasons, such as clinical diagnosis, treatment planning, and disease monitoring. However, volumetric segmentation remains computationally intensive due to the high dimensionality of medical imaging data and the anatomical variability across modalities such as CT and MRI~\cite{hatamizadeh2022unetr}.Traditional architectures, including 3D U-Net and its variants, have demonstrated strong performance~\cite{ronneberger2015u,cciccek20163d} but are often hindered by high memory consumption and inefficiency in real-time applications~\cite{isensee2021nnu}. Similarly, transformer-based models, including UNETR~\cite{hatamizadeh2022unetr}, leverage long-range dependencies but come with substantial computational overhead, hindering their deployment in resource-constrained environments~\cite{hatamizadeh2021swin}.

To address these limitations, we introduce {MLRU++}, a Multiscale Lightweight Residual UNETR++ architecture that combines the benefits of multiscale learning, lightweight residual design, and attention mechanisms to provide accurate yet efficient 3D segmentation. 
At the heart of MLRU++ is the Lightweight Convolutional Block Attention Module (LCBAM), which replaces conventional attention schemes with a streamlined alternative. While the standard CBAM~\cite{woo2018cbam} effectively improves feature quality by applying sequential channel and spatial attention, it still incurs non-trivial overhead due to multi-layer perceptrons and convolution operations. In contrast, LCBAM preserves the core benefits of dual attention while significantly reducing parameter count, making it well-suited for high-resolution 3D medical data. Through this hybrid of multiscale feature fusion and efficient attention, MLRU++ achieves strong segmentation performance across diverse medical datasets. By leveraging lightweight channel-spatial attention, MLRU++ achieves strong segmentation performance without incurring the computational burden typically associated with volumetric models. 

Our contributions are summarized as follows:
\begin{itemize}
    \item We propose a {\em lightweight residual UNETR++ backbone} that reduces parameter count while preserving representational capacity through residual connections. 
     \item We introduce a {\em LCBAM} with multiscale feature handling which fuses channel and spatial attention to enhance multi-resolution features across the encoder-decoder pathway and adaptively highlights informative features across scales.
     \item We validate MLRU++ across four large-scale datasets, showing that MLRU++ outperforms existing state-of-the-art models in accuracy, efficiency, and generalization with significantly reduced model complexity. 
\end{itemize}

\section{Related Work}
Medical image segmentation has been heavily influenced by U-Net~\cite{ronneberger2015u}, whose encoder-decoder architecture with skip connections has become foundational for subsequent models~\cite{zhou2018unet++,isensee2018nnu,cao2021swinunet}. As the focus shifted from 2D to volumetric segmentation, various extensions emerged to adapt to the 3D nature of clinical imaging data. Early 3D approaches handled volumetric inputs as stacks of 2D slices~\cite{milletari2016v,cciccek20163d}, Milletari et al.~\cite{milletari2016v} proposing a V-Net that processes entire 3D volumes using residual learning. Similarly, Cicek et al.~\cite{cciccek20163d} extended U-Net with 3D convolutions to improve segmentation from sparsely annotated volumes. Isensee et al.~\cite{isensee2021nnu} introduced nnUNet, a self-configuring pipeline that adapts its architecture and training strategy based on dataset characteristics, while Roth et al.~\cite{roth2018multi} applied multi-scale 3D fully convolutional networks for robust multi-organ segmentation. Later, enriching convolutional neural networks (CNNs) with broader contextual awareness led to the use of image pyramids~\cite{he2016deep}, large kernels~\cite{peng2017large}, dilated convolutions~\cite{yu2017dilated}, and deformable convolutions~\cite{dai2017deformable}. MedNeXt~\cite{roy2022mednext} introduced hierarchical feature processing and residual pathways with adaptive kernels, enhancing performance at the cost of higher computational complexity.

Following the success of Vision Transformers (ViTs) in capturing global dependencies~\cite{dosovitskiy2020image,cao2021swinunet}, several efforts have extended their application to medical image segmentation. Karimi et al.~\cite{karimi2021convolution} adapted ViTs for 3D segmentation by employing patch embeddings, while SwinUNet~\cite{cao2021swinunet} introduced shifted window mechanisms to efficiently integrate local and global features.

Building on these foundations, hybrid architectures that combine CNNs with transformers have gained traction due to their ability to balance local spatial detail with global semantic understanding. TransUNet~\cite{chen2021transunet} exemplifies this approach by combining CNN-extracted features with transformer-encoded global representations. TransFuse~\cite{wang2022transfuse} introduces a BiFusion module for effective modality fusion, and CoTr~\cite{xie2021cotr} further enhances efficiency by integrating deformable attention within a CNN backbone, reducing computational load while preserving spatial awareness. Fully transformer-based designs have also emerged. UNETR~\cite{hatamizadeh2022unetr} incorporates transformer encoders throughout a U-shaped architecture, allowing global context modeling directly from input-level patch tokens, and nnFormer~\cite{zhou2021nnformer} integrates hierarchical attention within a U-Net backbone. Most recently, Mosformer~\cite{huang2025mosformermomentumencoderbasedinterslice} advances this direction by introducing a momentum-based attention mechanism to capture inter-slice continuity in volumetric data.
In parallel, lightweight attention mechanisms have been explored to enhance feature representation with minimal computational overhead. One such method is the Convolutional Block Attention Module (CBAM)~\cite{woo2018cbam}, which sequentially applies channel and spatial attention to refine feature maps by focusing on the most informative regions. CBAM has demonstrated effectiveness across various vision tasks and can be integrated into transformer-based models to further improve segmentation accuracy while maintaining efficiency. In further advancement~\cite{10526382} introduced the Efficient Paired-Attention (EPA) block in UNETR++, a task-specific lightweight attention design tailored for 3D medical image segmentation. EPA performs spatial and channel attention in parallel using shared query and key projections. By learning complementary features through separate value paths and fusing the outputs with convolutional layers, EPA effectively captures enriched spatial-channel representations while maintaining computational efficiency. Recent lightweight hybrid approaches have aimed to balance efficiency with segmentation performance, DAE-Former~\cite{azad2023dae}, which incorporates dual attention mechanisms in a lightweight transformer pipeline; EMCAD~\cite{rahman2024emcad}, which enhances context fusion via multi-scale feature extraction. While effective in 2D tasks, these methods~\cite{azad2023dae,rahman2024emcad} are not well-suited for full 3D volumetric segmentation, as they often rely on 2D operations or attention schemes that do not scale effectively in 3D. Consequently, they struggle to capture the spatial continuity and contextual depth required in 3D medical image analysis. Models like UNETR++, and nnFormer~\cite{10526382,zhou2021nnformer} deliver state-of-the-art accuracy but may struggle with efficiency and generalization in real-world clinical settings. Addressing the trade-off between segmentation quality and computational efficiency remains an ongoing challenge in the field.

\begin{figure*}
\centering
    \includegraphics[width=0.8\linewidth]{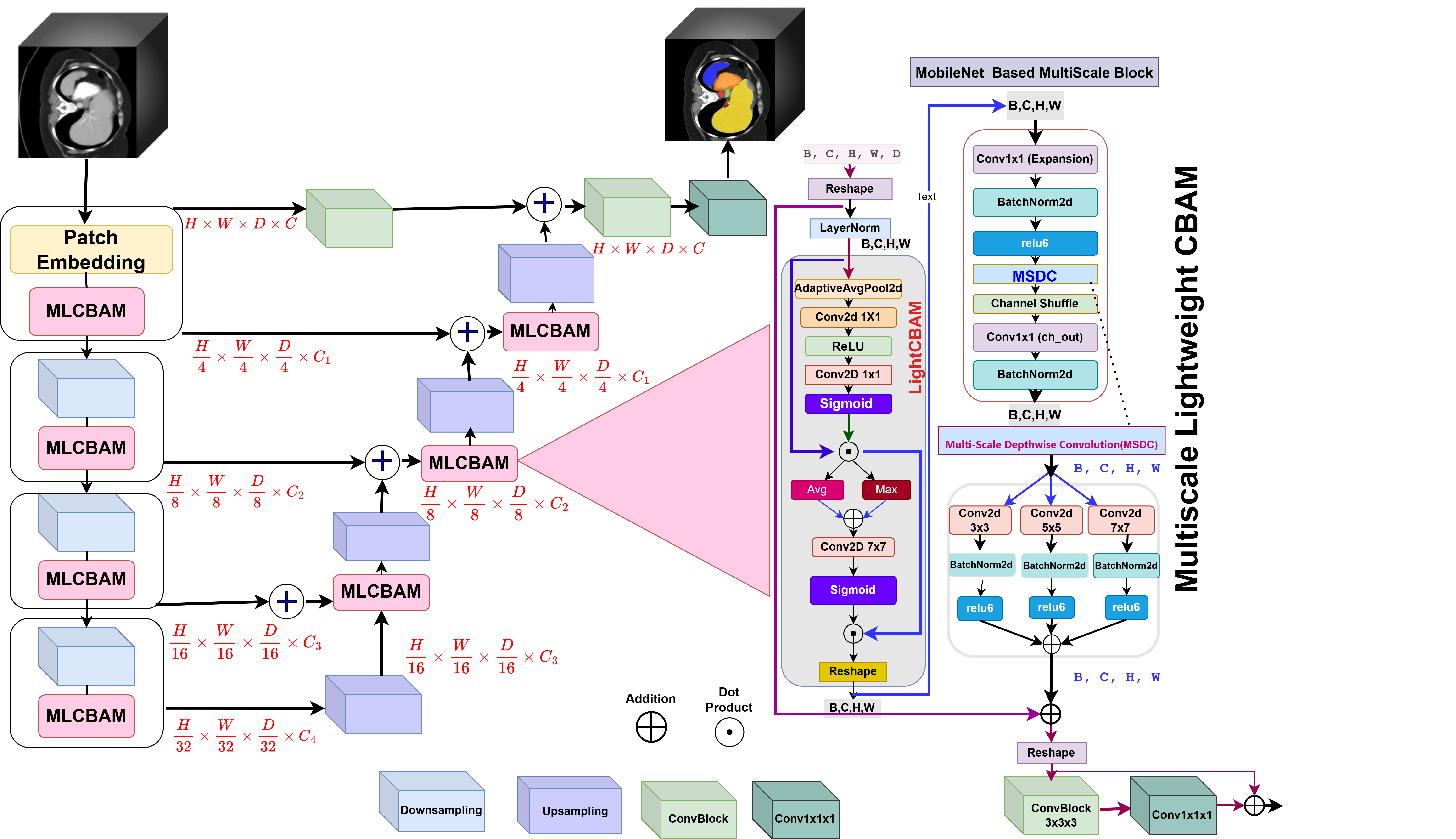}
    \caption{Overview of the proposed MLRU++ architecture. The network consists of a four-stage encoder–decoder design. Each encoder stage integrates the Lightweight Channel and Spatial Attention Block (LCBAM) for efficient feature extraction, while the decoder incorporates both LCBAM and the Multiscale Modulation Block (M\textsuperscript{2}B) to capture multiscale contextual information. Skip connections are used to preserve spatial detail, and deep supervision is applied at multiple decoder stages to guide learning at different resolutions.}
    \label{network}
\end{figure*}
\section{MLRU++ Architecture}

In this section, we present \textbf{MLRU++}, a novel architecture built upon a hierarchical encoder-decoder design specifically tailored for volumetric medical image segmentation. MLRU++ adopts a UNETR++-style encoder-decoder backbone enhanced with lightweight residual blocks, multiscale skip connections, and attention mechanisms. An overview of the architecture is illustrated in Fig.~\ref{network}.

Inspired by the UNETR++ framework~\cite{10526382}, MLRU++ introduces several key architectural enhancements: (i) a lightweight and efficient attention module, (ii) multiscale feature aggregation, and (iii) a MobileNet-based multiscale convolution block to improve both contextual richness and computational efficiency. The architecture consists of four encoder stages and four corresponding decoder stages, connected via skip connections. Each encoder stage performs downsampling and processes the input volume via our efficient Multiscale LCBAM (MLCBAM) attention module, progressively reducing spatial dimensions while extracting rich, multiscale features.

MLRU++ begins with a {\em patch embedding} module, which partitions the 3D input volume $x \in \mathbb{R}^{H \times W \times D}$ into a grid of non-overlapping patches of size $(P_1, P_2, P_3)$, producing $N = \left(\frac{H}{P_1} \times \frac{W}{P_2} \times \frac{D}{P_3}\right)$ patches, each reshaped as $x_u \in \mathbb{R}^{N \times (P_1, P_2, P_3)}$. Each patch is then linearly projected into a shared embedding space of dimension $C$, producing feature maps of shape $\frac{H}{P_1} \times \frac{W}{P_2} \times \frac{D}{P_3} \times C$. 

Following patch embedding, each encoder stage applies strided convolutions for spatial downsampling, followed by a {\em TransformerBlock} that integrates the proposed {\em Lightweight Convolutional Block Attention Module (LCBAM)}. This module captures both global and local dependencies via efficient channel and spatial attention mechanisms. Residual convolutional blocks in each stage further improve local feature encoding. The channel depth increases progressively across stages $[C_1, C_2, C_3, C_4]$ to encode hierarchical semantics.
The decoder consists of four upsampling stages that progressively restore spatial resolution, with skip connections from corresponding encoder stages. Each upsampling stage incorporates our MobileNet-based Multiscale Block (M\textsuperscript{2}B) to fuse multi-resolution context. The final segmentation output is produced by a ConvBlock that generates voxel-wise semantic predictions.


\subsection{Efficient Lightweight CBAM (LCBAM)}
 To improve computational efficiency while retaining the effectiveness of channel and spatial attention, we propose the LCBAM. Unlike the original CBAM, which relies on MLPs and full convolutions, LCBAM employs $1 \times 1$ convolutions, depthwise convolutions, and adaptive pooling to reduce parameters and FLOPs, making it suitable for deployment in constrained environments.
\subsubsection{Channel Attention}
The channel attention module avoids expensive MLPs by employing two lightweight $1 \times 1$ convolutions, combined with adaptive average pooling (AAP) and batch normalization (BN). Given an input feature map $M \in \mathbb{R}^{C \times H \times W}$, channel attention is defined as:
\begin{equation}
\alpha = \sigma \left( W_2 , \delta \left( \operatorname{BN} \left( W_1 , \text{AAP}(M) \right) \right) \right),
\label{eq:alpha}
\end{equation}
where $W_1 \in \mathbb{R}^{\frac{C}{r} \times C}$ and $W_2 \in \mathbb{R}^{C \times \frac{C}{r}}$ are $1 \times 1$ point-wise convolutions with a reduction ratio $r$. $\delta(\cdot)$ denotes the ReLU activation, and $\sigma(\cdot)$ the sigmoid function. Since these operations are independent of the spatial dimensions, the parameter cost is only $2C^2 / r$, significantly lower than CBAM's MLP-based alternative.

\subsubsection{Spatial Attention} 
For spatial attention, LCBAM improves over CBAM by replacing full convolutions with a {\em depthwise 7$\times$7 convolution}. The spatial attention map is computed as:
\begin{equation}
\beta = \sigma \left( f^{7 \times 7} \left[ \operatorname{Avg}(M_c) ; \operatorname{Max}(M_c) \right] \right),
\label{eq:beta}
\end{equation}
where $f^{7 \times 7}$ denotes a depthwise convolution with a $7 \times 7$ kernel, applied on the concatenation of average and max pooled features. As depthwise filters operate per channel, the number of learnable weights is fixed (49), resulting in a cost that scales with spatial resolution but not with channel depth.



\subsubsection{Attention Composition}
The final attention-enhanced feature map in LCBAM is computed by sequentially applying channel and spatial attention to the input feature map. Given an input $M \in \mathbb{R}^{C \times H \times W}$, the output $\hat{M}$ is obtained as:
\begin{equation}
\hat{M} = \beta \cdot (\alpha \cdot M),
\end{equation}
where $\alpha \in \mathbb{R}^{C \times 1 \times 1}$ is the channel attention map (Eq.~\ref{eq:alpha}), and $\beta \in \mathbb{R}^{1 \times H \times W}$ is the spatial attention map (Eq.~\ref{eq:beta}). The intermediate result $\alpha \cdot M$ represents the channel-refined feature, which is then modulated by the spatial attention mask $\beta$ to yield the final output.

This two-stage gating strategy enables the model to capture both \emph{global semantic dependencies} via channel attention and \emph{local structural context} via spatial attention, while maintaining a low computational overhead. The resulting attention-refined feature $\hat{M}$ is subsequently fed into the Transformer encoder block for further processing.




\subsection{MobileNet-Based Multiscale Block (\texorpdfstring{M\textsuperscript{2}B}{M2B})}

To enrich contextual representation between encoder and decoder stages while maintaining computational efficiency, we introduce {\em M\textsuperscript{2}B}. Inspired by the inverted residual design of MobileNetV2~\cite{Sandler_2018_CVPR}, M\textsuperscript{2}B adopts a \emph{pointwise $\rightarrow$ multiscale depthwise $\rightarrow$ pointwise} pattern tailored for efficient 3D volumetric segmentation. We extend this structure with two key enhancements:

\begin{enumerate}
    \item {\em Parallel Multiscale Depthwise Convolutions (MSDC)}: Multiple depthwise convolution branches with different kernel sizes (e.g., $3 \times 3$, $5 \times 5$, $7 \times 7$, etc.) are applied in parallel to capture multi-scale spatial features. Their outputs are summed to preserve both fine and coarse contextual information.

    \item {\em Channel Shuffle Operation}: As depthwise convolutions process each channel independently, a channel shuffle operation is applied before the projection layer to promote inter-channel feature interaction with minimal cost.
\end{enumerate}

The overall structure of M\textsuperscript{2}B can be expressed as a sequential operation: \[
\mathrm{PW}_{\text{proj}} \circ \mathrm{BN} \circ \mathrm{ReLU6} \circ  \operatorname{MSDC} \circ \mathrm{Add} \circ \mathrm{Shuffle} \circ \mathrm{PW}_{\text{exp}} \circ \mathrm{BN},
\]
where $\mathrm{PW}_{\text{exp}}$ and $\mathrm{PW}_{\text{proj}}$ denote the $1 \times 1 \times 1$ pointwise convolutions used for expansion and projection, respectively. $\mathrm{BN}$ and $\mathrm{ReLU6}$ denote batch normalization and the ReLU6 activation function. The $\operatorname{MSDC}$ operator performs parallel depthwise convolutions with multiple receptive fields and is formally defined in Section~3.3.

This design ensures that the M\textsuperscript{2}B block retains the lightweight nature of the original inverted residual block while significantly improving multi-scale feature representation. Importantly, the computational complexity and parameter count remain $\mathcal{O}(C)$, making it ideal for high-resolution 3D medical image segmentation tasks where efficiency and accuracy are both critical.

\subsection{Multi-scale Depthwise Convolution Block (MSDC)}
MSDC sub-block performs multiscale spatial filtering using parallel depthwise convolutions:

\begin{equation}
\operatorname{MSDC}(x)
   =\sum_{k\in\mathcal{K}}
     \underbrace{%
       \operatorname{R6}\!\bigl(\operatorname{BN}(\operatorname{DW}_{k}(x))\bigr)
     }_{\,\text{DWCB}_{k}(x)}
\label{eq:msdc},
\end{equation}
where $\mathcal{K} = {3, 5, 7}$ represents the kernel sizes used in each parallel branch. Each $\operatorname{DW}_{k}$ is a depthwise convolution followed by BatchNorm and ReLU6 activation. The summed result is added back to the original feature map to facilitate residual learning. This structure adds only $\mathcal{O}(C)$ parameters and remains independent of spatial resolution.






\section{Experiments}
\label{sec:experiments}

\subsection{Datasets and Evaluation Metrics}
Four publicly available benchmark datasets are used to evaluate the performance of the proposed \textbf{MLRU++} architecture in volumetric medical image segmentation tasks:

\begin{enumerate}
    \item {\em Synapse Multi-organ Dataset}~\cite{igelsias2015miccai} is composed of abdominal CT scans from 30 subjects annotated for 8 abdominal organs, \textit{i.e.,} spleen, right/left kidneys, gallbladder, liver, stomach, aorta, and pancreas.
    
    \item {\em BTCV Multi-organ Dataset}~\cite{landman2015miccai} contains 30 annotated abdominal CT volumes (24 for training, 6 for testing), covering 13 organs including adrenal glands, veins, and esophagus in addition to those in Synapse.
    
    \item {\em ACDC Cardiac Dataset}~\cite{bernard2018deep} is composed of short-axis cardiac MRIs from 100 patients with ground truth labels for right ventricle (RV), left ventricle (LV), and myocardium (MYO).
    
    \item {\em Decathlon Lung Dataset}~\cite{simpson2019large} is composed of 63 CT volumes, designed for binary segmentation of lung tumors versus background.
\end{enumerate}

We use Dice Similarity Coefficient (DSC)~\eqref{eq:DSC} across all datasets, and for Synapse, also report the 95\textsuperscript{th} percentile {Hausdorff Distance (HD95)} \eqref{eq:HD95} for boundary-level precision. They are mathematically expressed as follows:
\begin{equation}
\text{DSC}(Y, P) = \frac{2 \cdot |Y \cap P|}{|Y| + |P|},
\label{eq:DSC}
\end{equation}
where $Y$ and $P$ denote the sets of ground truth and predicted voxels; and 
\begin{equation}
\text{HD95}(Y, P) = \max \left\{ d_{95}(Y, P),\ d_{95}(P, Y) \right\},
\label{eq:HD95}
\end{equation}
where $d_{95}(\cdot, \cdot)$ computes the 95th percentile distance from one boundary to the other.

\subsection{Implementation Details}
{MLRU++} uses a four-stage encoder with channel sizes of 32, 64, 128, and 256. Each stage incorporates three LCBAM blocks for multiscale attention. The decoder mirrors the encoder with four upsampling stages, each starting with transposed convolution followed by LCBAM modules. In high-resolution settings (e.g., $128 \times 128 \times 64 \times 16$ for Synapse), the final decoder stage uses a standard $3\times3\times3$ convolution for efficiency.

Final predictions are produced using fused convolutional features passed through $3\times3\times3$ and $1\times1\times1$ convolutions. For fair comparison, all training configurations follow the setup in~\cite{zhou2023nnformer}.

Training is conducted on a single NVIDIA Tesla V100 GPU (32\,GB) with dataset-specific patch sizes:
\begin{itemize}
    \item {\em Synapse:} $64 \times 128 \times 128$, 
    \item {\em Lung:} $32 \times 192 \times 192$ and
    \item {\em ACDC:} $16 \times 160 \times 160$.
\end{itemize}

Deep supervision is used at all decoder levels with auxiliary loss weights $[0.57,\ 0.29,\ 0.14]$. The model is optimized with SGD (momentum = 0.99, Nesterov = True), learning rate = 0.01, and weight decay = $3 \times 10^{-5}$. 3D data augmentation includes elastic deformation, rotation, gamma correction, mirroring, and scaling.

\subsection{Loss Function}
To guide learning, we adopt a composite loss combining soft Dice loss and cross-entropy loss to balance overlap quality and voxel-wise accuracy:
\begin{equation}
\scalebox{0.85}{$
L_{\text{seg}}(Y, P) = 1 - \sum_{i=1}^{I} \left( \frac{2 \sum_{v=1}^{V} Y_{v,i} P_{v,i}}{\sum_{v=1}^{V} (Y_{v,i}^2 + P_{v,i}^2)} \right)
- \sum_{v=1}^{V} \sum_{i=1}^{I} Y_{v,i} \log(P_{v,i})
$},
\end{equation}
where \(I\) is the number of classes, \(V\) the number of voxels, and \(Y_{v,i}, P_{v,i}\) are the ground truth and predicted probability for class \(i\) at voxel \(v\).

\begin{table*}[tbp]
\centering
\caption{State-of-the-art comparison on the Synapse abdominal multi-organ segmentation dataset. The table reports Dice Similarity Coefficient (DSC, \%) and 95th-percentile Hausdorff distance (HD95, mm) for each organ, along with the model complexity measured by the number of parameters (in millions). \textbf{Bold} values indicate the best performance. Our proposed MLRU++ achieves superior segmentation accuracy while maintaining competitive model complexity.
}
\resizebox{\textwidth}{!}{%
\small
\begin{tabular}{lccccccccccccc}
\toprule
Methods & Spleen & Kidney (right) & Kidney (left) &Liver & Gallbladder & Aotra & Pancreas & Stomach & \multicolumn{2}{c}{Average} \\
 & & & & & & & & & DSC $\uparrow$  & HD95 $\downarrow$   & Params $\downarrow$  \\
\midrule
U-Net \cite{ronneberger2015u} & 86.67 & 68.60 & 77.77 & 93.43 & 69.72 & 89.07 & 53.98 & 75.58 & 76.85  & 39.70  & 14.8 M\\
nnUNet \cite{isensee2021nnu} & 91.68 &  88.46 & 84.68 & \textbf{97.13} &  78.82 &  93.04 & \textbf{83.23} & 83.34 & 87.33  & 10.91  & 19.07 M\\

TransUNet2D\cite{chen2021transunet} & 85.08 & 77.02 & 81.87 & 94.08 & 63.16 & 87.23 & 55.86 & 75.62 & 77.49  & 31.69  &96.07 M \\
Swin-UNet\cite{cao2021swinunet} & 90.66 & 79.61 & 83.28 & 94.29 & 66.53 & 85.47 & 56.58 & 76.60 & 79.13  & 21.55  &27.17 M \\
UNETR\cite{hatamizadeh2022unetr}   & 85.00 & 84.52 & 85.60 & 94.57 & 56.30 & 89.80 & 60.47 & 70.46 & 78.35  & 18.59  &92.58 M \\
MISSFormer \cite{huang2022missformer} & 91.92 & 82.00 & 85.21 & 94.41 & 68.65 & 86.99 & 65.67 & 80.81 & 81.96 & 18.20  & 42.46 M \\
DAE-Former\cite{azad2023dae} & 91.82 & 82.39 &87.66  &95.08  &71.65  &87.84  &63.93  & 80.77 &82.63   & 16.39  & 48.01 M \\

CoTr \cite{xie2021cotr} & 94.93 & 86.80 & 87.67 & 96.37 & 62.90 & 92.43 & 78.84 & 80.46 & 85.05  & 9.04 &  46.51 M \\

nnFormer \cite{zhou2023nnformer} & 90.51 & 86.25 & 86.57 & 96.84 & 70.17 & 92.04 & 82.41 & 86.83 & 86.57  & 10.63  & 149.12 M \\
{MOSformer\cite{huang2025mosformermomentumencoderbasedinterslice}} &92.29   &83.58  &\textbf{90.32}  &95.96  &71.90   &88.95   & 74.14 &87.87  &85.63   & 13.40  &77.00 M \\ 
{UNETR++\cite{10526382}} &\textbf{95.77} & {87.18} & {87.54} & 96.42 & 71.25 & {92.52} & 81.10 & 86.01 & {87.22}  & {7.53}  & 42.96 M\\ \hline
MLRU++ (Encoder LCBAM only) & {89.90} & {87.48} & {88.18} & 97.06 &69.75 &92.21   & {83.89} & 82.37 & 86.36 & {10.92}   & 44.92 M \\ 
\textbf{MLRU++} & {91.77} & \textbf{87.60} &  {87.67} & {96.86}  &\textbf{73.11} &\textbf{93.29} & {82.39} &\textbf{87.88}  &\textbf{87.57} & \textbf{7.53}   &{46.09 M}  \\
\bottomrule
\end{tabular}}
\label{tab:synapse}
\end{table*}


\section{Results and Analysis}
\label{sec:results}

The proposed {MLRU++}, incorporating LCBAM modules in both encoder and decoder stages, consistently outperforms state-of-the-art results across four different medical segmentation benchmarks. DSC is used as the primary metric for all datasets, while the Synapse dataset also reports HD95.

\subsection{Synapse Dataset Performance} 
\begin{figure}[tbp]
    \centering
    \includegraphics[width=\linewidth]{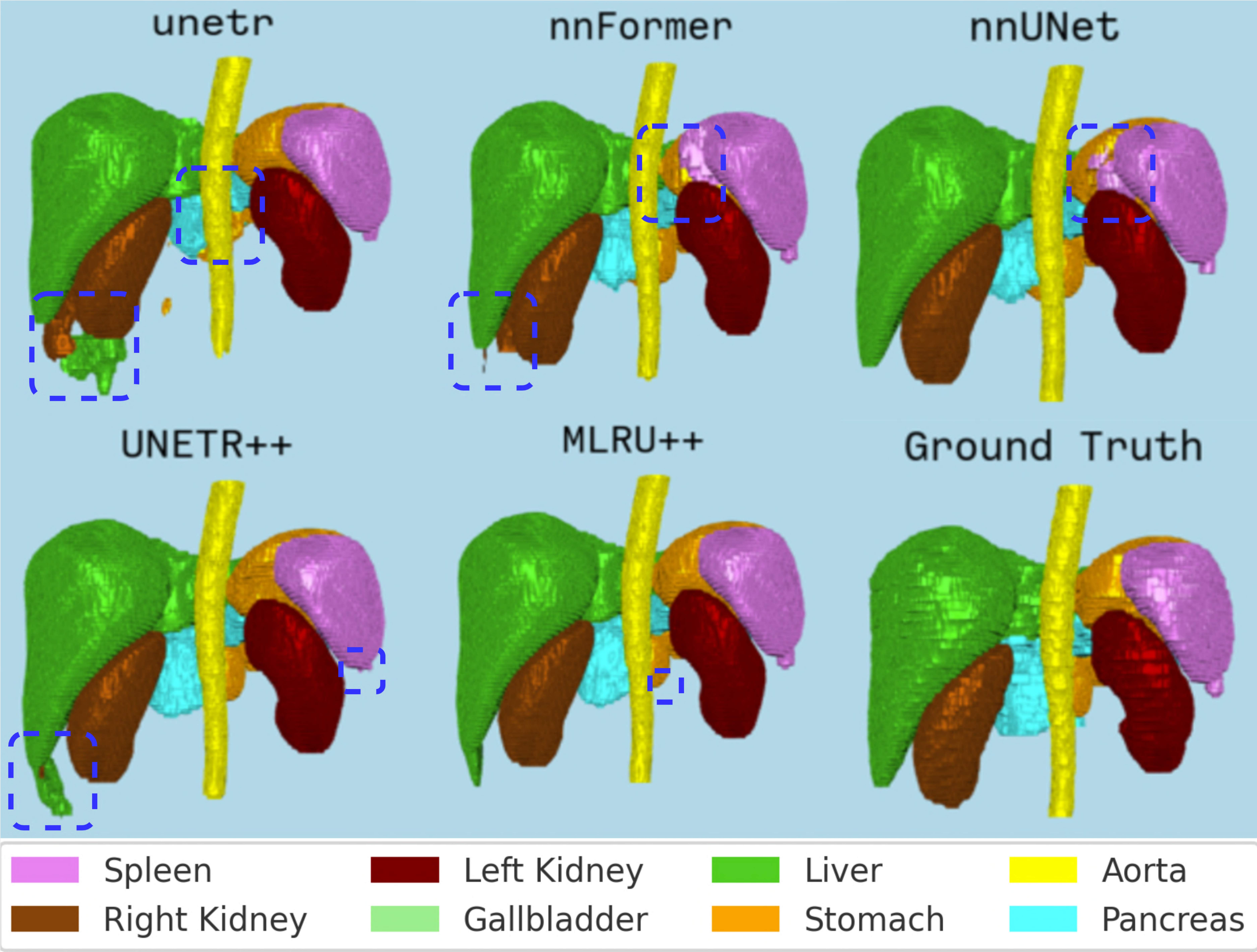}
    \caption{Baseline methods vs MLRU++ on the Synapse dataset: Blue dashed boxes highlight inaccurate segmentation.}
    \label{fig:base}
\end{figure}

\begin{figure}[tbp]
\includegraphics[width=\linewidth]{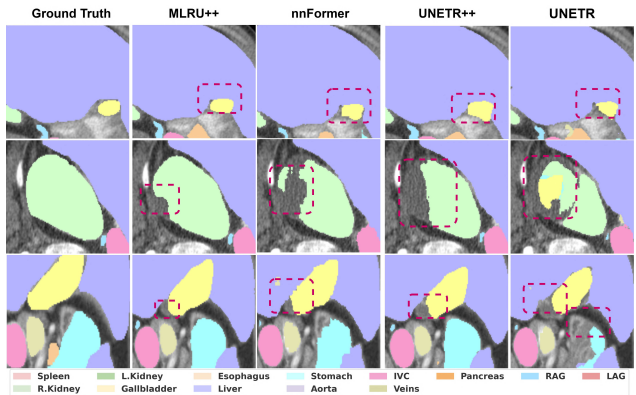}
    \caption{Qualitative comparison between baseline methods and the proposed MLRU++ on the Synapse dataset. Regions with inaccurate segmentation are highlighted using purple dashed boxes.
    }
    \label{fig:enter-syn}
\end{figure}

As shown in Table~\ref{tab:synapse}, {MLRU++} achieves the highest average DSC of \textbf{87.57\%} and the lowest HD95 of {7.53}, outperforming all baselines. Compared to UNETR++, MLRU++ improves performance across most organs, notably:
\begin{itemize}
    \item Gallbladder: 71.25 $\rightarrow$ {73.11};
    \item Aorta: 92.52 $\rightarrow$ {93.29};
    \item Right Kidney: 87.18 $\rightarrow$ {87.60}; and 
    \item Left Kidney: 87.54 $\rightarrow$ {87.67}.
\end{itemize}
Additionally, compared to the widely adopted nnUNet~\cite{isensee2021nnu}, MLRU++ offers measurable improvement (87.33 vs. 87.57) while maintaining a modular, efficient design. A comparative analysis with baseline models, as illustrated in Fig.~\ref{fig:base}, demonstrates the effectiveness of our approach. Qualitative results (visualizations) are provided in Fig.~\ref{fig:enter-syn}.

\subsection{BTCV Dataset Performance}
\begin{figure}[tbp]
\centering
\includegraphics[width=0.94\linewidth]{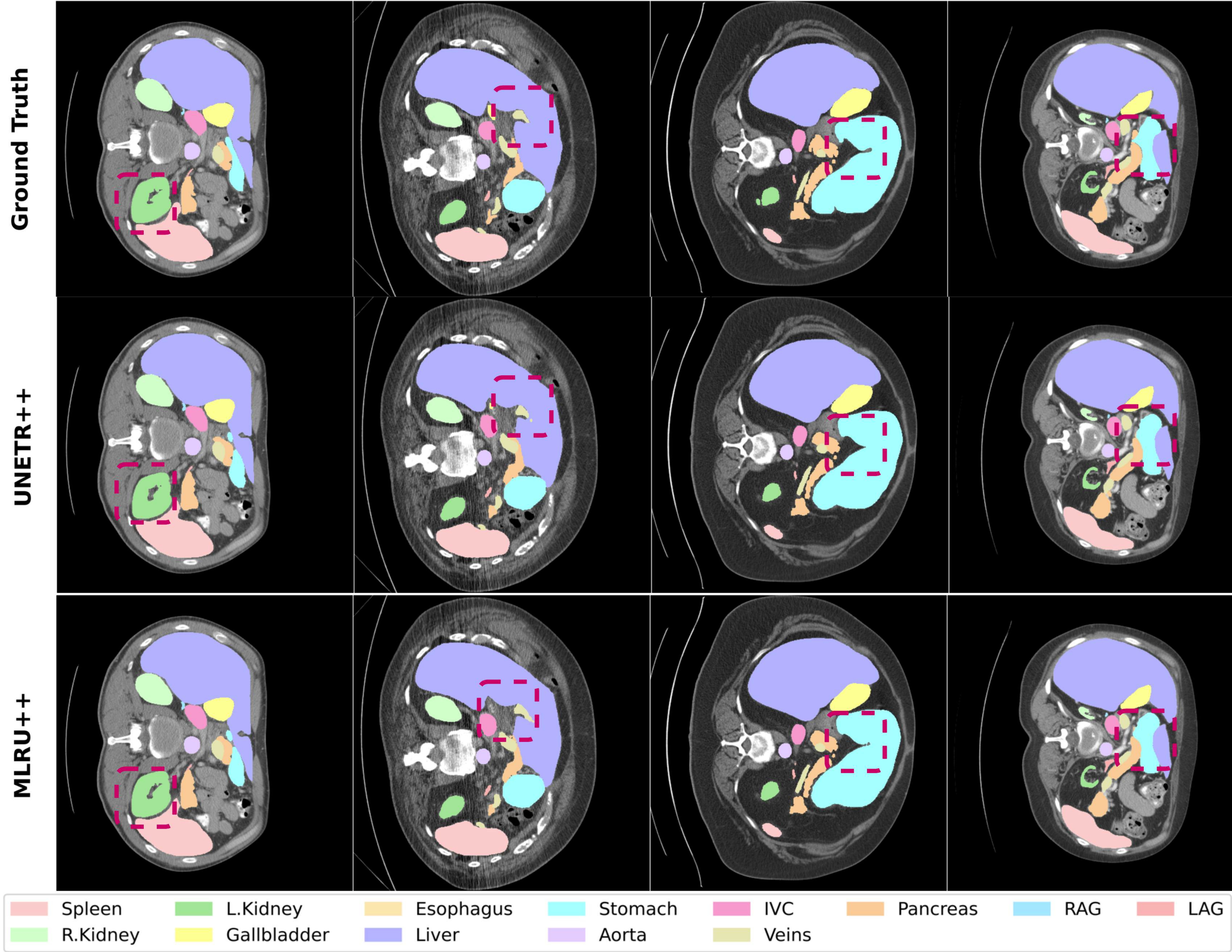}
    \caption{Qualitative comparison of segmentation results on the BTCV multi-organ dataset. From top to bottom: input image with ground truth, predictions from UNETR++, and the proposed MLRU++. MLRU++ shows enhanced delineation of both large and small anatomical structures, demonstrating superior spatial consistency and boundary precision, especially for challenging organs such as the adrenal glands and pancreas.
    }
    \label{fig:enter-btvc}
\end{figure}

Table~\ref{tab:BTVC} presents results on the BTCV multi-organ abdominal dataset. MLRU++ achieves the best overall DSC of \textbf{85.20\%}, surpassing transformer-based models like UNETR++, TransUNet, and nnFormer, as well as classical architectures like nnUNet. Significant improvements are seen in smaller, complex structures:
\begin{itemize}
    \item Left Adrenal Gland (LAG): 74.29;
    \item Inferior Vena Cava (IVC): 88.34;
    \item Right Kidney: 94.96; and
    \item Left Kidney: 94.93.
\end{itemize}
Compared to UNETR++ (85.04) and nnUNet (83.16), MLRU++ shows consistent and reliable gains. Visual segmentation comparisons are shown in Fig.~\ref{fig:enter-btvc}.

\begin{table*}[tbp]
\centering
\caption{Quantitative comparison of segmentation performance on the BTCV multi-organ dataset. We report Dice Similarity Coefficient (DSC, \%) for each organ and the average across all organs. MLRU++ achieves the highest average DSC, demonstrating its effectiveness over existing CNN and transformer-based models. \textbf{Bold} indicates the best scores.\textbf{Note:} RAG and LAG refer to the Right Adrenal Gland and Left Adrenal Gland, respectively. 
} 
\resizebox{\textwidth}{!}{\small
\begin{tabular}{lcccccccccccccc}
\toprule
\textbf{Method} & \textbf{Spleen} & \textbf{RightKid} & \textbf{LeftKid} & \textbf{Gall} & \textbf{Esoph} & \textbf{Liver} & \textbf{Stomach} & \textbf{Aorta} & \textbf{IVC} & \textbf{Veins} & \textbf{Pancreas} & \textbf{RAG} & \textbf{LAG} & \textbf{Mean} \\
\midrule
nnUNet  \cite{isensee2021nnu}     & 95.95 & 88.35 & 93.02 & \textbf{70.13} & 76.72 & 96.51 &{ 86.79} & 88.93 & 82.89 &{ 78.51} &{ 79.60} & \textbf{73.26} & 68.35 & 83.16 \\
CoTr\cite{xie2021cotr} &95.80 &92.10 &93.60 &70.00 & 76.40 &96.30 &85.40 & \textbf{92.00} &83.8 0 & 78.70 &77.50 &69.40  &66.50 &82.88 \\

TransUNet\cite{chen2021transunet}  &95.20 & 92.70 & 92.20  &66.20  &75.70  & 96.90  & 88.90 & \textbf{92.00} & 83.30  & 79.10  & 77.50  & 69.60  &66.60  & 82.76  \\
nnFormer  \cite{zhou2023nnformer}   & 94.58 & 88.62 & 93.68 & 65.29 & 76.22 & 96.17 & 83.59 & 89.09 & 80.80 & 75.97 & 77.87 & 70.20 & 66.05 & 81.62 \\
UNETR++ \cite{10526382}   & 95.61 & 94.94 & 94.92 &65.38 & \textbf{78.60} & 97.07 & 87.50 & 91.41& 88.23 & 79.02 & 86.70 &  73.06 & 73.08 & 85.04 \\ 
 \hline
\textbf{MLRU++ (Ours)}   &\textbf{96.44} &\textbf{94.96}&\textbf{94.93} & 65.45&{75.47}&\textbf{97.16}&\textbf{90.00}&{90.38}&\textbf{88.34}& \textbf{79.71}& \textbf{87.31}&{73.10}&\textbf{74.29} &\textbf{85.20} \\
\bottomrule
\end{tabular}}
\label{tab:BTVC}
\end{table*}

\subsection{Decathlon Lung Dataset Performance}
MLRU++ is evaluated on the Decathlon-Lung dataset to assess generalizability. As summarized in Table~\ref{tab:lung}, it achieves the highest Dice score of {81.12\%}, surpassing strong baselines such as UNETR++ and nnFormer.

\begin{table}[tbp]
\centering
\small
\caption{Dice Similarity Coefficient (DSC \%) on the Decathlon Lung dataset across various state-of-the-art models. MLRU++ achieves the highest DSC, indicating superior lung segmentation performance. \textbf{Bold} highlights the best result.
}
\begin{tabular}{lc}
\toprule
\textbf{Model} & \textbf{DSC (\%)} \\
\midrule
UNETR~\cite{hatamizadeh2022unetr} & 73.29 \\
nnUNet~\cite{isensee2021nnu} & 74.31 \\
SwinUNETR & 75.55 \\
nnFormer~\cite{zhou2023nnformer} & 77.95 \\
UNETR++~\cite{10526382} & 80.68 \\
 \hline
\textbf{MLRU++ (Ours)} & \textbf{81.12} \\
\bottomrule
\end{tabular}
\label{tab:lung}
\end{table}

\subsection{ACDC Cardiac Dataset Performance}
\begin{figure*}[tbp]
    \centering
\includegraphics[width=\linewidth]{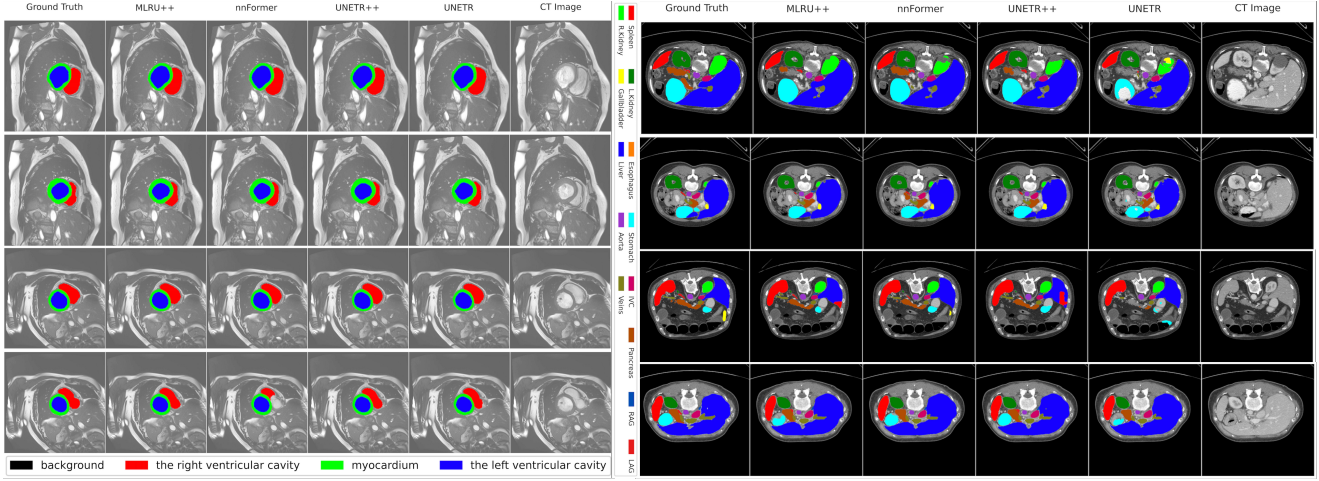}
    \caption{Qualitative segmentation results comparing MLRU++ with other methods on two datasets: Synapse (left) and ACDC (right). Ground truth contours are overlaid on input images, followed by predictions from UNETR, nnUNet, UNETR++, and the proposed MLRU++. MLRU++ demonstrates superior anatomical boundary adherence and clearer segmentation, particularly for small and complex structures across both abdominal and cardiac regions.
    }
    \label{fig:syn-acdc}
\end{figure*}

We further evaluate MLRU++ on the ACDC dataset, which includes annotations for RV, LV, and myocardium. As shown in Table~\ref{tab:ACDC}, MLRU++ achieves the highest average DSC of {93.00\%}, outperforming hybrid CNN-Transformer models. These results indicate MLRU++'s robustness in segmenting anatomically varied and temporally dynamic structures. Qualitative results (visualizations0 are provided in Fig.~\ref{fig:syn-acdc}.

\begin{table}[tbp]
\centering
\caption{
Performance comparison on the ACDC cardiac segmentation dataset. MLRU++ achieves the highest average Dice score across right ventricle (RV), left ventricle (LV), and myocardium (MYO), demonstrating superior cardiac structure segmentation. \textbf{Bold} indicates the best scores.}
\resizebox{0.45\textwidth}{!}{
\begin{tabular}{lcccc}
\hline
\textbf{Methods} & \textbf{RV} & \textbf{Myo} & \textbf{LV} & \textbf{Avg} \\
\hline
TransUNet \cite{chen2021transunet}      & 88.86 & 84.54 & 95.73 & 89.71 \\
Swin-UNet \cite{cao2021swinunet}        & 88.55 & 85.62 & 95.83 & 90.00 \\
UNETR \cite{hatamizadeh2022unetr}      & 85.29 & 86.52 & 94.02 & 88.61 \\
MISSFormer\cite{huang2022missformer}  & 86.36 & 85.75 & 91.59 & 87.90 \\
nnUNet \cite{isensee2021nnu}           & 90.96 & 90.34 & 95.92 & 92.41 \\
nnFormer\cite{zhou2023nnformer}      & 90.94 & 89.58 & 95.65 & 92.06 \\
{UNETR++\cite{10526382}} & 91.89 & 90.61 & 96.00 & 92.83 \\
{MOSformer\cite{huang2025mosformermomentumencoderbasedinterslice}}& 90.86 & 89.65 & 96.05 & 92.19 \\
{EMCAD\cite{rahman2024emcad}} & 90.65 & 89.68 & 96.02 & 92.12 \\
 \hline
\textbf{MLRU++ without M\textsuperscript{2}B } &  {91.30}  &  {90.53}  &   {95.96}  &  {92.59}  \\
\textbf{MLRU++} & \textbf{91.96}  & \textbf{90.83}  &  \textbf{96.20}  & \textbf{93.00}  \\
\hline
\end{tabular}
\label{tab:ACDC}
}
\end{table}

MLRU++ demonstrates strong generalization across organ types and imaging modalities, showing particular effectiveness on challenging, small-scale anatomical structures due to its multiscale attention design and deep supervision strategy.

\section{Ablation Studies}
Our goal is to examine the impact of the LCBAM module and the multiscale feature extraction block (denoted as M\textsuperscript{2}B), in MLRU++, on segmentation performance on two different datasets: Synapse and ACDC datasets.

\subsection{Effectiveness of Multiscale LCBAM Modules}
Table~\ref{tab:synapse} demonstrated that incorporating both LCBAM and M\textsuperscript{2}B modules yield the highest Dice score (87.57\%) on the Synapse dataset. To further validate the individual contribution of these components, we compare different model variants, summarized in Table~\ref{tab:epa-ablation}.

We begin by analyzing the impact of using only LCBAM modules (without M\textsuperscript{2}B). When LCBAM is applied solely in the encoder (second last row), the performance is slightly improved over the UNETR baseline, but remains below the full MLRU++ model. The variant using LCBAM in both encoder and decoder (without M\textsuperscript{2}B) performs better (86.66\%), yet still falls short of the proposed architecture that includes the multiscale block. This performance gap clearly highlights the importance of the M\textsuperscript{2}B block in enabling the decoder to effectively handle multiscale semantic features. Moreover, when compared to other strong baselines such as UNETR++~\cite{10526382} and EMCAD~\cite{rahman2024emcad}, MLRU++ not only improves accuracy but also maintains competitive model complexity and FLOPs.

\begin{table}[tbp]
\centering
\caption{
Ablation study evaluating the impact of the Multiscale LCBAM (M\textsuperscript{2}B) module in MLRU++ on the Synapse dataset. Results demonstrate that integrating the multiscale block significantly improves segmentation performance while maintaining efficient model complexity.}
\resizebox{1\linewidth}{!}{%
\begin{tabular}{lccc}
\toprule
\textbf{Model Variant} & \textbf{Params (M)} & \textbf{FLOPs (G)} & \textbf{DSC (\%)} \\
\midrule
UNETR (Baseline)~\cite{hatamizadeh2022unetr} & 92.49 & 75.76 & 78.35 \\
UNETR++~\cite{10526382} & \textbf{42.96} & \textbf{47.98} & 87.22 \\

MLRU++ (LCBAM Only & 44.60 & 63.63 & 86.36 \\
Encoder + Decoder) &&&\\
MLRU++ (LCBAM  & 44.92 & 64.81 & 86.66 \\
Only in Encoder) &&&\\
 \hline
 \textbf{MLRU++ (LCBAM + M\textsuperscript{2}B)} & 46.09 & 66.06 & \textbf{87.85} \\
\bottomrule
\end{tabular}}
\label{tab:epa-ablation}
\end{table}

\subsection{Generalization to ACDC Dataset}
As before, a similar trend is observed in the ACDC dataset (see Table~\ref{tab:ACDC}), where MLRU++ with both LCBAM and M\textsuperscript{2}B modules outperforms its LCBAM-only variant. This demonstrates that the multiscale block plays a critical role across different medical imaging tasks, including both abdominal and cardiac segmentation.

\subsection{Discussion}
These results confirm that while LCBAM enables attention to important spatial and channel-wise features, it is the M\textsuperscript{2}B block that provides strong multiscale feature aggregation and contextual learning in the decoder. Their combination leads to the best performance and highlights the effectiveness of our full MLRU++ architecture.


\section{Conclusion}

In this work, we introduced MLRU++, a novel and efficient architecture for volumetric medical image segmentation that combined lightweight attention mechanisms with multiscale feature integration. Our design incorporated the LCBAM (Lightweight Channel and Bottleneck Attention Module) in both the encoder and decoder, and introduced a multiscale feature enhancement block (M\textsuperscript{2}B) in the decoder to address spatially complex structures across scales. Through extensive experiments on four diverse and widely used datasets: Synapse, BTCV, ACDC, and Decathlon-Lung, we demonstrated that MLRU++ achieved state-of-the-art performance while maintaining competitive computational efficiency. We consistently outperformed existing methods and baselines, including nnUNet, UNETR++, and nnFormer, particularly on anatomically small or heterogeneous structures. Ablation studies further validated the importance of both LCBAM and the multiscale decoder module, confirming that their combination yielded significant performance improvements. The consistent performance gains across datasets underscored the generalizability and robustness of MLRU++.

\balance
{
    \small
 
    \bibliographystyle{splncs04}  

    \bibliography{main}
}

\end{document}


\appendix
\section*{\centering Supplementary Material}
\section*{Appendix A: Computational Efficiency of LCBAM vs. CBAM and ViT Attention}

\subsection*{A.1 Channel Attention Analysis}

CBAM~\cite{woo2018cbam} applies two fully connected (MLP) layers to global descriptors from average and max pooling. The formulation is:
\begin{equation}
\mathcal{M}_c(F) = \sigma\left( W_1 \cdot \text{ReLU}(W_0 F_{\text{avg}}) + W_1 \cdot \text{ReLU}(W_0 F_{\text{max}}) \right)
\end{equation}
where $W_0 \in \mathbb{R}^{C/r \times C}$, $W_1 \in \mathbb{R}^{C \times C/r}$, and $r$ is the reduction ratio (typically 16).
\textbf{Total parameters for CBAM channel attention:}
$
\text{Params}_{\text{CBAM}} = 2 \cdot \left(C \cdot \frac{C}{r} + \frac{C}{r} \cdot C \right) = \frac{4C^2}{r}
$

\subsection*{A.2 LCBAM Channel Attention}

LCBAM replaces the MLP with lightweight $1 \times 1$ convolution layers:

\begin{equation}
\mathcal{M}_c(F) = \sigma\left( \text{Conv}_{1\times1} \left( \text{ReLU} \left( \text{Conv}_{1\times1} (\text{AvgPool}(F)) \right) \right) \right)
\end{equation}

\textbf{Total parameters for LCBAM:}
\[
\text{Params}_{\text{LCBAM}} = 2 \cdot \left(C \cdot \frac{C}{r} \right) = \frac{2C^2}{r}
\]

This yields a \textbf{50\% reduction} in parameters for the channel attention module compared to CBAM.

\subsection*{A.3 Spatial Attention Comparison}

CBAM applies a $7\times7$ convolution over concatenated pooling layers:
\begin{equation}
\mathcal{M}_s(F) = \sigma\left( f^{7 \times 7} \left[ \text{AvgPool}(F); \text{MaxPool}(F) \right] \right)
\end{equation}
\textbf{FLOPs (CBAM Spatial):}
$
\text{FLOPs}_{\text{CBAM}} = 7 \cdot 7 \cdot 2 \cdot H \cdot W = 98 \cdot H \cdot W $
\section*{LCBAM Spatial Simplification.} LCBAM employs either depthwise $3 \times 3$ or $1 \times 1$ convolutions:
$
\text{FLOPs}_{\text{LCBAM}} = 3 \cdot 3 \cdot 2 \cdot H \cdot W = 18 \cdot H \cdot W \quad (\text{or lower with } 1 \times 1)
$

\subsection*{A.4 Summary: CBAM vs. LCBAM}

\begin{table}[h]
\centering
\caption{Computational Comparison Between CBAM and LCBAM (Ours)}
\resizebox{0.9\linewidth}{!}{%
\begin{tabular}{@{}lccc@{}}
\toprule
\textbf{Module} & \textbf{CBAM} & \textbf{LCBAM} & \textbf{Reduction} \\
\midrule
Channel Attention & $\frac{4C^2}{r}$ & $\frac{2C^2}{r}$ & 50\% fewer params \\
Spatial Attention & $98HW$ & $9HW$ & $\sim$90\% fewer FLOPs \\
\textbf{Total} & Moderate & Ultra-lightweight & Significant \\
\bottomrule
\end{tabular}%
}
\label{tab:cbam_lcbam}
\end{table}

\vspace{2mm}
\noindent
\textbf{Conclusion:} LCBAM drastically reduces computational burden while preserving attention effectiveness—ideal for lightweight 3D applications and real-time inference.

\section*{Appendix B: Proof of Computational Efficiency in Multiscale LCBAM}

\subsection*{B.1 Parameter Count Comparison}

Let the input be $B \times C \times H \times W$ and tokens $N = H \cdot W$.
\paragraph{CBAM:}
$
P_{\text{CBAM}} = 2 \cdot \frac{C^2}{r} + 98
$
\paragraph{ViT-style Self-Attention:}
$
P_{\text{ViT}} = 4C^2
$

\paragraph{Multiscale LCBAM (ours):}
$
P_{\text{LCBAM}} = 2 \cdot \frac{C^2}{r} + 49C
$
$
P_{\text{MSCB}} = C^2 + 3C \quad \text{(for } k \in \{1,3,5\})
$
$
P_{\text{Total}} = \mathcal{O}(C^2 + C)
$

\subsection*{B.2 Empirical FLOPs Estimate ($C=64$, $H=W=32$)}
\begin{table}[h]
\centering
\caption{Empirical Cost Comparison}
\begin{tabular}{@{}lcc@{}}
\toprule
\textbf{Module} & \textbf{Parameters} & \textbf{FLOPs} \\
\midrule
ViT Attention & $\sim$16K & $\sim$270M \\
CBAM & $\sim$0.6K & $\sim$1.2M \\
LCBAM (ours) & $\sim$12K & $\sim$0.3M \\
Multiscale LCBAM (ours) & $\sim$16K & $\sim$3.5M \\
\bottomrule
\end{tabular}
\end{table}

\subsection*{B.3 Theoretical Justification of Efficiency}

\begin{itemize}
  \item \textbf{No Quadratic Complexity:} LCBAM avoids $O(N^2C)$ attention cost present in ViT.
  \item \textbf{Depthwise Convolution Usage:} MSCB’s depthwise separable convolutions yield $\sim\frac{1}{C}$ the cost of full convolutions.
  \item \textbf{Reduced Attention Overhead:} Attention uses $1\times1$ convs and global pooling, no MLPs or dense FC layers.
\end{itemize}

\subsection*{Conclusion}

Our Multiscale LCBAM achieves superior trade-offs in efficiency and scalability, yielding nearly an order-of-magnitude reduction in parameters and FLOPs over ViT-style attention and CBAM. This efficiency makes it particularly well-suited for 3D segmentation, edge inference, and low-power deployments.
\bibliographystyle{ieeenat_fullname}
\bibliography{main}